\documentclass[12pt]{iopart}
\usepackage{graphicx}
\usepackage{bbm}
\usepackage{iopams}
\def\be{\begin{equation}}
\def\ee#1{\label{#1}\end{equation}}
\def\ba{\begin{eqnarray}}
\def\ea#1{\label{#1}\end{eqnarray}}
\def\a{\alpha}
\def\d{\delta}
\def\h{\eta}
\def\k{\kappa}
\def\l{\lambda}

\def\p{\pi}
\def\r{\rho}
\def\t{\tau}
\def\f{\varphi}
\def\om{\omega}
\def\bfe{\bi{e}}
\def\bv{\bi{v}}
\def\bu{\bi{u}}
\def\mbS{\mathbf{{\mathcal S}}}
\def\mbH{\mathbf{{\mathcal H}}}
\def\bM{\bi{M}}
\def\bH{\bi{H}}
\def\bbf{\bi{f}}
\def\bF{\bi{F}}
\def\bR{\bi{R}}
\def\br{\bi{r}}
\def\bx{\bi{x}}
\def\bX{\bi{X}}
\def\bbs{\boldsymbol\sigma}
\def\bbx{\boldsymbol\xi}
\def\bfo{\boldsymbol{\omega}}
\begin{document}
\title[Chiral transport]{Transport of flexible chiral objects in a uniform shear flow}

\author{Peter Talkner, Gert-Ludwig Ingold, and Peter H\"anggi}
\address{Institut f\"ur Physik, Universit\"at  Augsburg,
Universit\"atsstra{\ss}e 1, D-86135 Augsburg, Germany}
\ead{peter.talkner@physik.uni-augsburg.de}
\date{\today}

\begin{abstract}
The transport of slightly deformable chiral objects in a uniform shear flow is
investigated. Depending on the equilibrium configuration one finds up to four
different asymptotic states that can be distinguished by a lateral drift
velocity of their center of mass, a rotational motion about the center
of mass and deformations of the object. These deformations influence the
magnitudes of the principal axes of the second moment tensor of the considered
object and also modify a scalar index characterizing its chirality. Moreover,
the deformations induced by the shear flow are essential for the phenomenon of
dynamical symmetry breaking: Objects that are achiral under equilibrium
conditions may dynamically acquire chirality and consequently experience a
drift in the lateral direction.
\end{abstract}

\pacs{83.50.-v, 05.60.Cd, 47.85.Np, 83.10.Pp}
%\keywords{particle transport in fluids, shear flow, chirality, hydrodynamic interactions, symmetry breaking }
\submitto{\NJP}

\maketitle

\section{Introduction}
\label{I}
The idea of achieving enantiomer separation  on the basis of specific transport
properties of chiral objects in flowing fluids \cite{KSTH,HM,CHR} is very
attractive because it opens the perspective to tackle this also industrially
relevant task by purely physical means without the need of specific chemical
agents. 

Howard, Lightfoot and Hirschfelder \cite{HLH} were the first to suggest the use
of transport properties of asymmetric objects in order to separate differently
crystallized enantiomers.  Later, de Gennes discussed a mechanical way of
separating chiral crystals \cite{dG}. Only recently, a lateral drift of
screw-like objects in shear flows predicted by Brenner \cite{B} was
experimentally demonstrated for bodies with a linear extension of the order of
centimeters \cite{CC}, millimeters \cite{MAD} and micrometers 
%\cite{MFPS}. The direction of this drift, resulting from the coupling of
\cite{MFPS}. The direction of this drift, resulting from the coupling of
the rotational and translational motion of a chiral object in a fluid, differs
for objects with opposite chirality. The mutual interactions between  flowing
fluids  and immersed chiral objects has been the subject of various
recent experimental studies.  The generation of flow patterns by
actively rotating chiral objects was investigated in Ref.~\cite{GW}. The
control of actively moving nano-structured propellers was studied in
Ref.~\cite{GF}. In Ref.~\cite{DRLP} the formation of chiral assemblies made 
of achiral components in a vortex flow is described.

Theoretical studies of the influence of the chirality of a macroscopic object
on its transport properties were performed for sedimentation \cite{DM} and for
shear flows \cite{MD}. Chiral planar three-atomic molecules were studied in
clover leave vortex flows \cite{KSTH} and in flows in channels of various
geometries \cite{E1,E2}. In these two-dimensional studies fluctuating forces
were taken into account in order to model the action of thermal fluctuations
exerted on the molecules by the surrounding fluid. All studies
\cite{KSTH,DM,MD,E1,E2} though are restricted to rigid objects.

Watari and Larson considered three-dimensional, four-atomic molecules which, in
equilibrium, assume the form of a  regular tetrahedron, which as such is
achiral with respect to geometry \cite{WL}. When put into a shear flow these
molecules are distorted in an asymmetric way due to differently deformable
bonds between the atoms. As a consequence these molecules experience a lateral
drift.

In the present work we study the motion of four-atomic molecules with {\it
different} equilibrium bond-lengths but identical bond-strengths, see in
eq.~(\ref {FF}) below,  being  suspended  in a uniform shear flow. The main
mechanisms leading to a lateral drift of the molecules, i.e. to a motion
perpendicular to the flow direction and to the direction in which the flow
speed changes, can be summarized as follows. The shear field of the flow leads
to a rotation of the molecule which then, like a small propeller, moves in, or
opposite to the direction of the vorticity of the flow. We demonstrate that the
rotation is essentially determined by the advection of the atoms with the fluid
motion and only little influenced by hydrodynamic interactions between the
atoms. In a uniform shear flow a coupling between rotational and translational
motion, however, only is possible due to hydrodynamic interactions between the
atoms.

As for the motion of molecular motors \cite{HM}, the direction of the lateral
drift is difficult to predict. In general it depends on all parameters of the
system and also on the initial conditions with which the molecule is started.
These initial conditions may belong to different domains of attraction,
eventually leading to different attractors, which may display different forms
of rotation of the molecule, and consequently different drift behaviors, even
with opposite signs.

Due to the deformability of the considered molecules, chiral attractors may
exist for molecules with achiral equilibrium configurations and consequently
also achiral molecules may experience a lateral drift displaying a dynamical
symmetry breaking. The lateral drift, however, vanishes on average because for
a molecule with achiral equilibrium configuration chiral attractors can only
come in mutually mirror symmetric pairs.

The paper is organized as follows. In Section~\ref{II} we introduce a class of
``spiral'' molecules, the geometry of which is described by only three
parameters, and specify their equations of motion in terms of bond strengths
and hydrodynamic interactions.  Further, we introduce the most important
quantities needed for the analysis of the numerical solutions of the equations
of motion (Section~\ref{IIA}). In Section~\ref{III} we present the results
concerning the center of mass motion (Section~\ref{IIIA}), the rotational
motion (Section~\ref{IIIB}) and the internal motion (Section~\ref{IIIC}) with
particular emphasis on the dynamical symmetry breaking (Section~\ref{IIIC1}).
The role of hydrodynamic interactions is discussed in Section~\ref{IV}. The
paper closes with a summary and an outlook in Section~\ref{V}.

\section{The model}
\label{II}
\subsection{Spiral molecules and equation of motion}
In the present case study we consider ``molecules'' consisting of four
identical ``atoms'' modeled by spheres of radius $a$. The equilibrium
configuration of a free molecule, i.e. a stable configuration with balanced
bonding forces between the atoms, is assumed to be of helical form. This form
is constructed by positioning the centers of the spheres on a cylinder mantle
of radius $\r$. Given the location of the first atom, $\bR_1$, the next one,
$\bR_2$, is found upon rotation by an angle $\f$ about the cylinder axis
combined with a translation by $h$ parallel to the cylinder axis, and so on
with the third and forth atom, yielding the respective positions $\bR_3$ and
$\bR_4$.  Hence, if the cylinder axis coincides with the $z$-direction of a
coordinate system  and the first atom lies on the $x$-axis of the coordinate
system, then the following vectors point to the centers of the atoms sitting on
a spiral:
\be
\bR_n = \r \bfe_x \cos [(n-1) \f] + \r \bfe_y \sin [(n-1) \f] + h (n-1) \bfe_z\:, \quad n = 1,2,3,4\:.
\ee{xn}

We model the bonding forces between the atoms of such a ``spiral'' molecule
by nonlinear, so-called FENE-Fraenkel (Finitely Extensible Nonlinear Elastic)
springs \cite{HJL}. If $\br$ is the vector connecting two atoms, then the force
acting between these atoms is given by
\be
\bbf= k (r-l)\frac{(r_+-l)(l-r_-)}{(r_+-r)(r-r_-)}\frac{\br}{r} \;, \quad \mathrm{for}\; r_-<r<r_+\:,\\
%r_{\pm}&=(1\pm s)l
\ee{FF}
where $r = |\br|$ is the distance between the atoms, $l$ the equilibrium length
of the spring, $k$ the spring-constant describing the Hookean behavior of the
spring for small extensions about the equilibrium length, and $s<1$ is a
nonlinearity parameter of the spring. At the extensions 
\be
r_{\pm} =(1\pm s)l
\ee{rpm} 
the force diverges so that the bond-length between two atoms is restricted to
the indicated, finite range.

In contrast to Ref.~\cite{WL}, we choose {\it identical} force parameters $k$
and $s$ for the bonds, only the equilibrium bond lengths $l$ may differ. Hence,
a possible chirality of the equilibrium configuration of a molecule is solely
determined by its geometry: According to the standard definition, a molecular
configuration is {\it achiral} if a movement of the molecule, i.e. a
combination of translations and rotations in three-dimensional space, exists such that it
matches its mirror image; in any other case the configuration is {\it chiral}.

When the molecule is suspended in a flowing fluid, advective forces act on the
individual atoms, which in addition mutually influence their motions by
hydrodynamic interactions. The total action of the fluid on the atoms will be
described by a mobility tensor $\bH_{n,m}$ relating the forces acting on the
$m$th  particle to the velocity of the $n$th particle.  The deterministic,
overdamped motion of the molecule in a fluid flow with the velocity field
$\bv(\br)$ is then given by
\be
\dot{\br}_n = \bv(\br_n) + \sum_{m=1}^4 \bH_{n,m} \cdot \bbf_m\:,
\ee{dem}
where $\bbf_m$ is the total force acting on the $m$th atom, i.e. the sum of the
FENE-Fraenkel forces (\ref{FF}) exerted by the other three atoms, and the dot
denotes differentiation with respect to time. For sufficiently small spheres
representing the atoms, inertial terms can be neglected \cite{Purcell}.  We
shall describe the mobility by the Rotne-Prager tensor which includes the
hydrodynamical interactions up to second order in the ratio $a/r_{n,m}$ of the
atomic radius $a$ and the distance $r_{n,m}$ between the $n$th and the $m$th
atom. It is given by \cite{RP}
\be
\bH_{n,m} = \frac{1}{6 \p \h a} \left[ \d_{n,m} \mathbbm{1} + (1-\d_{n,m})\frac{3 a}{4 r_{n,m}}\mbH_{n,m}\right] \;,
\ee{HH}
with
\be
\mbH_{n,m}  =  \left \{
  \begin{array}{ll}
\displaystyle \left (1+\frac{2 a^2}{3 r_{n,m}^2}\right ) \mathbbm{1} + \left (1-\frac{2 a^2}{r_{n,m}^2} \right ) \frac{\br_{n,m} \br_{n,m}}{r_{n,m}^2} \; & \mathrm{for}\;  r_{n,m} \geq 2 a\\*[3mm]
\displaystyle \left (1-\frac{9 r_{n,m}}{32 a}  \right )\mathbbm{1} + \frac{3 \br_{n,m} \br_{n,m}}{32 a r_{n,m}}& \mathrm{for}\;  r_{n,m}< 2a
  \end{array}
\right . \:,
\ee{RP}
where $\mathbbm{1}$ denotes the unit tensor in the three-dimensional space,
$\h$ the viscosity of the fluid, and $\br_{n,m}=\br_m - \br_n$ the vector
pointing from the position of the $n$th to that of the $m$th atom.

In passing we mention that thermal noise leads to additional random
contributions in the equations of motion, see Ref.~\cite{D}, yielding
\be
\dot{\br}_n = \bv(\br_n) + \sum_{m=1}^4 \bH_{n,m} \bbf_m + 2 k_B T \sum \bbs_{n,m} \cdot \bbx_m(t) \:,
\ee{rem}
where $\bbx_n(t)$ are mutually independent three-dimensional Gaussian, white
noise vectors, satisfying
\be
\langle \bbx_n(t) \bbx_m(s) \rangle = \d_{n,m} \mathbbm{1} \d(t-s)\: .
\ee{xx}
The fluctuation-dissipation theorem requires that the coupling to the noise,
$\bbs_{i,j}$, is related to the Rotne-Prager tensor according to
\be
\bH_{n,m} = \sum_{k=1}^4 \bbs_{n,k}\cdot \bbs_{j,m} \: .
\ee{Hss}

\begin{figure}
\begin{center}
\includegraphics[width=0.3\textwidth]{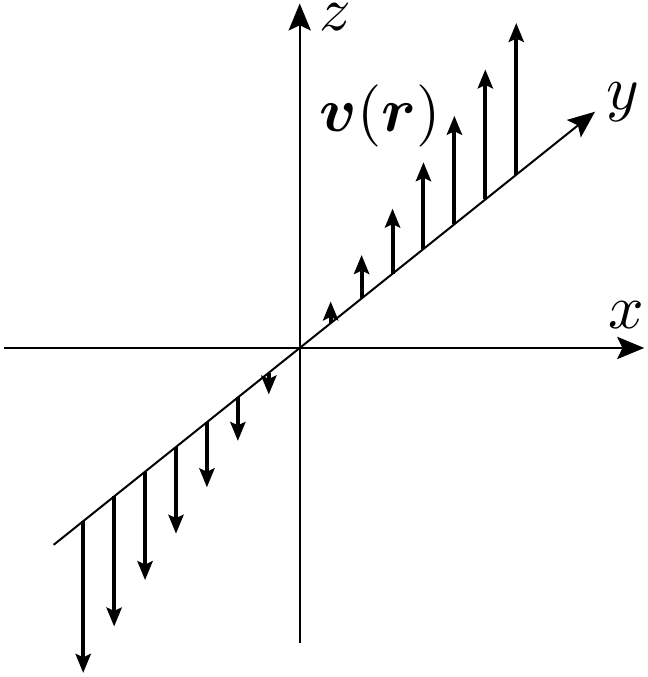}
\end{center}
\caption{Uniform shear flow: The velocity is independent of $x$, points in 
$z$ direction, and varies its magnitude proportionally to $y$. }
\label{fig1}
\end{figure}

In the present work though we neglect all influences stemming from thermal
noise, and, hence restrict ourselves to the discussion of the deterministic
equations of motion in (\ref{dem}).  We shall consider a uniform shear flow
given by
\be
v_x = v_y =0, \qquad v_z = \k y\:.
\ee{shv}
where $\k$ denotes the shear rate, see Figure~\ref{fig1}.  Even in this
seemingly simple case the deterministic equations of motion can only be solved
by numerical means. For that purpose we employed a LSODA code of variable order
and step-length \cite{RH,P} as available from the open source SciPy package
\cite{NP}.

In the numerical treatment we used dimensionless variables where lengths are
measured in units of the atomic radius $a$ and times in units of the inverse
shear rate $\k^{-1}$. Apart from six parameters which  fix the equilibrium
geometry of a molecule consisting of four atoms in general, or three in the
present case of a spiral molecule as defined by eq.~(\ref{xn}), two further parameters are
needed for the characterization of the forces. As such we will choose the
nonlinearity parameter $s$ and the dimensionless spring constant $k_f = k/(6 \p
\eta a^2 \k)$. With this choice the deterministic equations of motion read
\be
\dot{\bx}_n = \mbS \bx_n + k_f \sum_{m=1}^4 \mbH_{n,m} \cdot \bF_m \:,
\ee{del}
where $\mbS$ is the shear tensor, which has vanishing matrix elements apart
from the $z$-$y$ element, being  unity, and $\bF_m$ denotes the dimensionless
force exerted on the $j$-th particle by the adjacent particles, reading
\ba
\bF_m &= \sum_{n\neq m} \frac{x_{n,m} - \ell_{n,m}}{1-(1-x_{n,m}/\ell_{n,m})^2}\frac{\bx_{n,m}}{x_{n,m}}\\
&\hspace{4truecm}\mathrm{for}\; (1-s)\ell_{n,m} < x_{n,m} < (1+s) \ell_{n,m} \:.\nonumber
\ea{F}
Here $\bx_n=\br_n /a$, $\bx_{n,m} =\bx_{n} -\bx_m $, $x_{n,m}=|\bx_{n,m}| $ and
$\ell_{n,m} =|\bR_n -\bR_m|/a$ denote the dimensionless positions, distance
vectors, the actual distances and the equilibrium distances between atoms $n$ and
$m$, respectively.

Possible initial conditions for the integration of the equation of motion
(\ref{del}) are given by any set of atomic positions that obey the restrictions
imposed by the FENE-Fraenkel forces (\ref{F}). Here we confine ourselves to
equilibrium configurations of the free molecules. For further details of how
the orientations are chosen, see Sect.~\ref{III}.

\subsection{Characteristics of the mechanical molecular states}
\label{IIA}

In the framework of the overdamped dynamics of eq.~(\ref{del}) the mechanical
state of the molecule is uniquely described by the positions $\bx_n(t)$ of the
four atoms.  A reduced description is provided by the center of mass of the
molecule consisting of four  atoms of equal mass, consequently reading
\be
\bX = \frac{1}{4}\sum_n \bx_n\: .
\ee{Xt}
The spatial orientation of the molecule can conveniently be characterized by
the three principal axes of the symmetric tensor of second moments of the
atomic positions, $\bM$, defined by its elements
\be
M_{i,j} = \frac{1}{4} \sum_n (x^n_i -X_i)(x^n_j -X_j) \:,
\ee{M}
where $x^n_i$ denotes the $i$-th Cartesian position component of the $n$-th
atom. The orientations of the principal axes are determined by the normalized
eigenvectors $\bu_i$ and their magnitudes by the corresponding eigenvalues
$\l_i$ of the tensor of second moments. Hence, they result from the eigenvalue
problem
\be
\bM \bu_i = \l_i \bu_i\:,
\ee{evM}
where $\bM = (M_{i,j})$.

The velocity vector $ \boldsymbol{\omega}$ of the instantaneous rotation of the
molecule in the frame co-moving with the molecule's center of mass is
determined by the orientation and velocity of any body-fixed tripod. Here we
use the principal axes of the second moments to obtain
\be
\bfo= -\frac{1}{2} \sum_i\dot{\bu_i} \times \bu_i\:,
\ee{bom}
where the dot denotes a derivative with respect to time and $\bu \times \bv$
the vector product of the vectors $\bu$ and $\bv$. By means of the eigenvalue
equation (\ref{evM}) one may express $\dot{\bu_i}$ in terms of the time
derivative of $\bM$, to yield
\be
\bfo=\frac{1}{2} \sum_{i\neq j} \frac{\bu_j\cdot \dot{\bM} \bu_i }{\l_j -\l_i} \bu_j \times \bu_i\: ,
\ee{oM}
where $\bu \cdot \bv$ denotes the scalar product of the vectors $\bu$ and $\bv$.

As a measure of the deformation of a molecule we consider the ratio of volumes
taken in the distorted state and the equilibrium shape. As molecular volume one
may assign the square root of the product of the magnitudes of the principal
axes, i.e.,
\be
V = (\l_1 \l_2 \l_3)^{1/2}\:.
\ee{Vo}

Whether a molecule is chiral cannot be decided on the basis of its second
moments of positions. For this purpose, we use the isotropic chirality index
$G_0$, which is invariant under translations and rotations of the molecule
\cite{SLGN}, in its dilation invariant form. For a molecule consisting of four
atoms of equal mass and size it becomes
\be
G_0 = \frac{1}{3} \sum_{i,j,k,l}\frac{\bx_{i,j}\cdot(\bx_{k,l} \times \bx_{i,l})\: (\bx_{i,j} \cdot \bx_{j,k})\:(\bx_{j,k} \cdot \bx_{k,l})}{(x_{i,j} x_{j,k} x_{k,l})^2 x_{i,l}}\:.
\ee{G0}
The sum is extended over all sets $i,j,k,l$, taken from the permutations of
$\{1,2,3,4\}$.  This index has the general property that it assumes the same
absolute value but opposite signs for enantiomers, i.e. pairs of configurations
that are mirror images of each other. Hence, achiral configurations have a
vanishing chirality index.

\begin{figure}
\begin{center}
\includegraphics[width=.95\columnwidth]{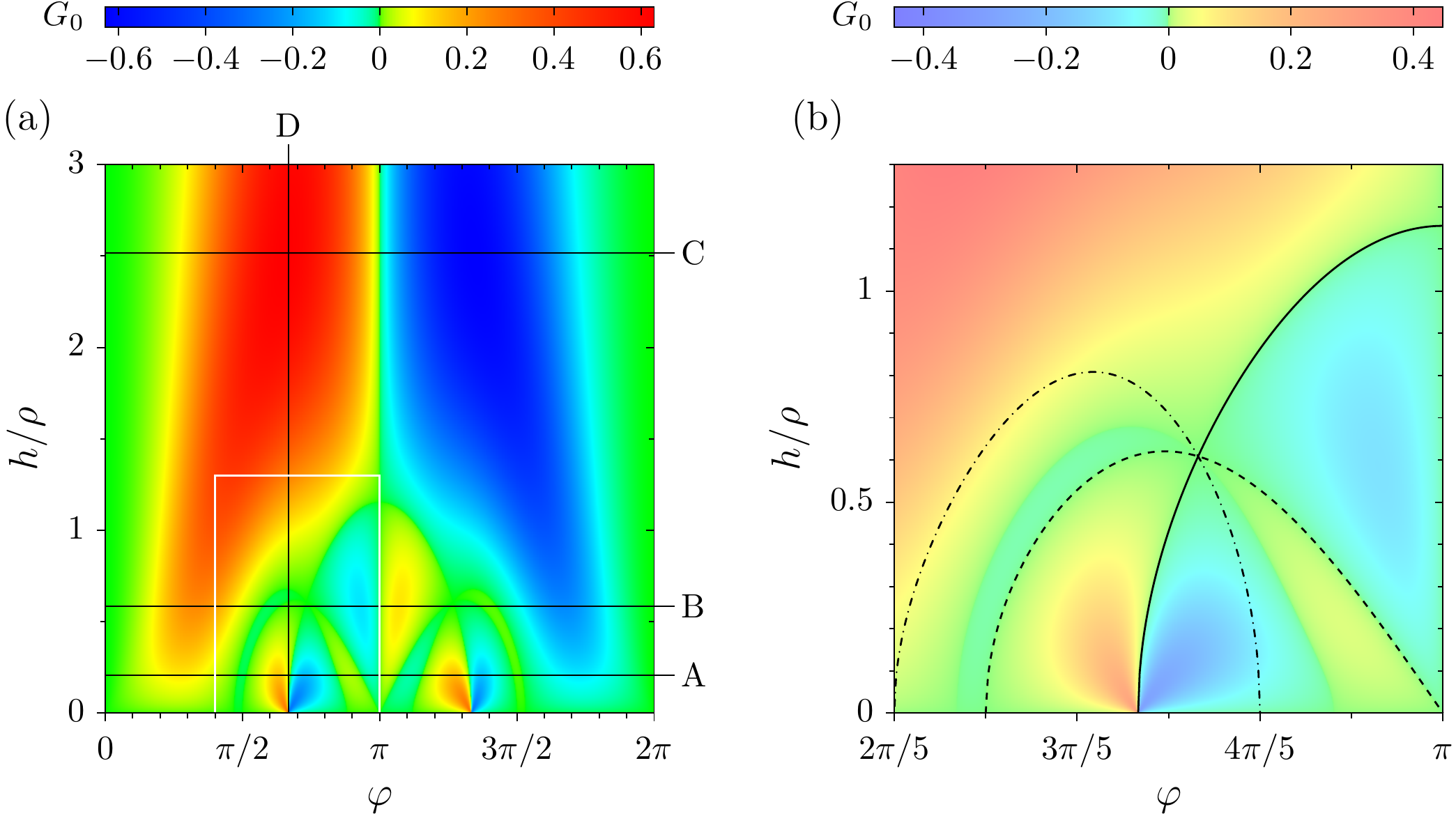}
\end{center}
\caption{The chirality index $G_0$ defined in eq.~(\ref{G0}) is displayed for
spiral molecules as a function of the screw angle $\f$ and the relative rise
$h/\r$, see eq.~(\ref{xn}). On the panel (a) the symmetry $G_0 \to -G_0$ upon a
reflection on the line $\f=\p$ is clearly visible. The four black lines labeled
by A -- D indicate four cuts through the parameter plane, see the text in
Section~\ref{III} as well as Fig.~\ref{fig5} below. The lines A and B are
defined by local maxima of the chirality index. On C and D the absolute maximum
of $G_0$ is located. The panel (b) represents a blow-up of the region marked
in the panel (a) by the white lines. Note that for the purpose of better
visibility different color scales are used in the two panels. The black lines in (b)
indicate those curves on which two of the three edge lengths $l(k)$, $k=1,2,3$
between atoms with positions $\bR_n$ and $\bR_{n+k}$ are equal, see the
Appendix. The solid line represents achiral configurations with $l(1)=l(2)$
which have five equally long edges.  It starts at the singularity of $G_0$ at
$\f =2 \p/3$ and $h/\r =0$ and ends at $\f =4 \p/3$ and $h/\r=0$.  Along the
dashed line $l(1)=l(3)$ holds. These configurations have two and four equally
long edges and are also achiral. On the dashed-dotted line, where  $l(2)=l(3)$,
the configurations possess two triples of equally long edges, and the chirality
index is non-zero everywhere except at the points with $h/\r=0$ and at the
crossing point of the three lines which corresponds to a regular tetrahedron.
Note that there is still a third curve recognizable by the green color on which
the chirality index vanishes in spite of the chirality of the corresponding
configurations.}
\label{fig2}
\end{figure}

For a spiral molecule, the chirality index is a function of the relative rise
$\r/h$ and the screw angle $\f$. It is an antisymmetric function about the line
$\f =\p$, see Fig.~\ref{fig2}(a), and possesses a singularity at $\f=2 \p/3$
and $h/\r =0$. Along the lines $h/\r \approx -(\f  -2 \p/3)$ for $\f < 2\p/3$,
and $h/\r \approx 0.95(\f -2 \p/3)$  for $\f > 2\p/3$, the chirality index
forms a ridge and a valley, respectively.  Approaching the singularity along
the ridge, the chirality index converges to the approximate value $0.289$ while
it approaches the opposite value $-0.289$ along the bottom of the valley.

\begin{figure}
\begin{center}
  \includegraphics[width=.3\columnwidth]{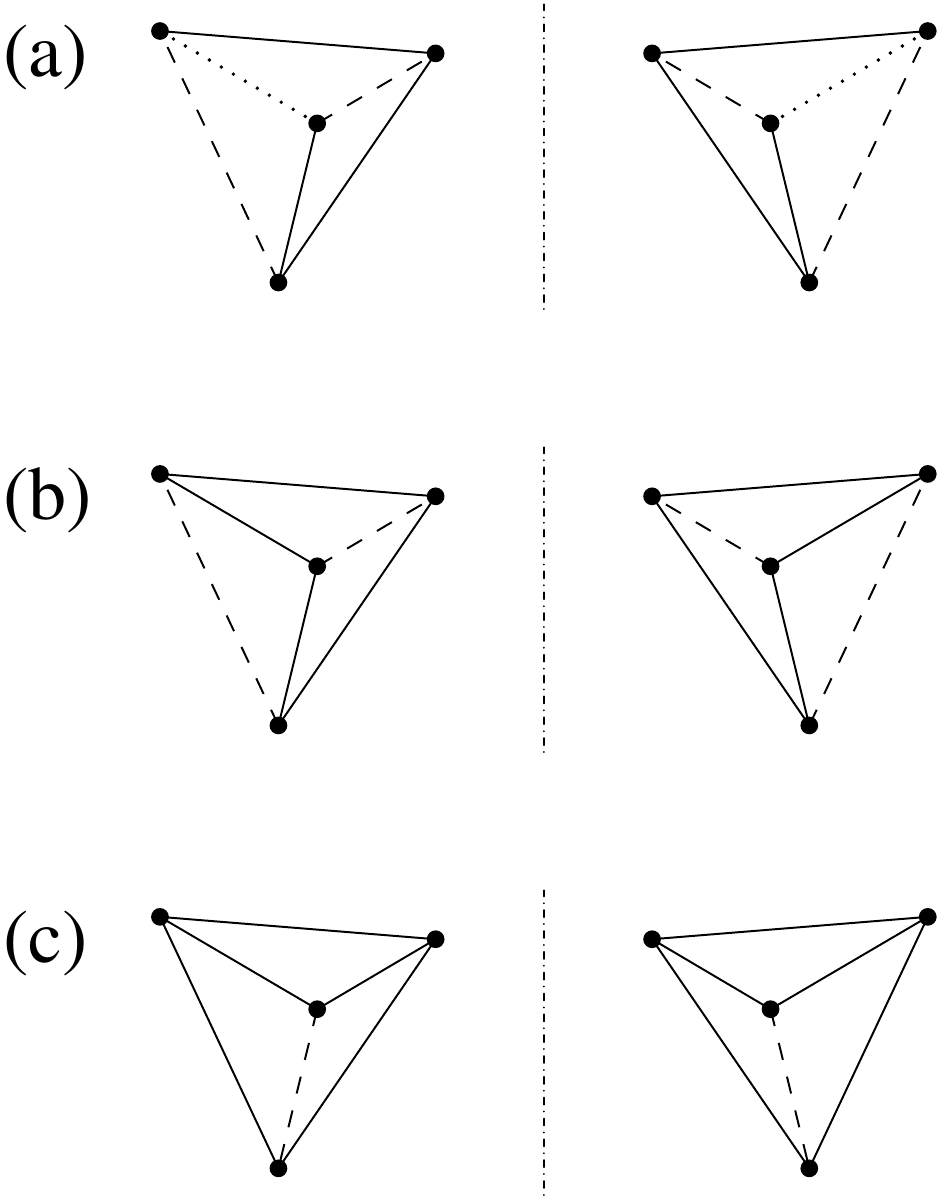}
\end{center}
\caption{Panel (a) illustrates a generic spiral molecule with three and two
edges of equal lengths and a sixth edge of different length. The edges of equal
lengths are  drawn correspondingly. Panel (b) presents a configuration on the
dashed curve displayed in Fig.~\ref{fig2}(b) with two and four equally long
edges while the configuration in panel (c) has five equally long edges
representing a configuration on the solid curve depicted in Fig.~\ref{fig2}(b).
The right part of the figure represents the corresponding mirror images. The
cases (b) and (c) are achiral.}
\label{fig3}
\end{figure}

Since all configurations with $h/\r =0$ are planar, and, hence achiral,  the
chirality index exactly vanishes on the line $h/\r =0$. In the remaining part
of the $\f$-$h/\r$-plane the chirality index has three extrema on each side of
the symmetry line $\f = \p$. Local extrema are located at $\f \approx 0.8908
\p$, $h/\r \approx 0.2073$ and $\f \approx 0.9291 \p$, $h/\r \approx 0.584$.
The absolute maximum is at $\f \approx 0.656 \p$, $h/\r \approx 2.516$, see
Fig.~\ref{fig2}(a). The chirality index vanishes on the solid and dashed curves
in Fig.~\ref{fig2}(b). These curves result from constraints on the bond
lengths (see Appendix).  Along the dashed curve emerging from $\f = \p/2$, $h/\r=0$, and
ending in $\f =\p$, $h/\r =0$, the configurations possess a group of four  and
another one of two equally long edges, see Fig.~\ref{fig3}(b).  Since the
corresponding configurations possess a symmetry plane, they  are  achiral. The
faces of these tetrahedra are made of identical isosceles triangles. Also the
configurations on the curve starting at $\f =2\p/3 $, $h/\r =0$ and reaching
$\f =\pi$, $h/\r = 2/\sqrt{3}$ are achiral; the faces of these configurations
consist in two identical equilateral and two identical isoscele triangles;
consequently five of the six edges have the same length, see
Fig.~\ref{fig3}(c). The analytic forms of the achiral curves are derived in the
Appendix.  Examples of different types of configurations are displayed in
Fig.~\ref{fig3}.  There exists a third node line of $G_0$ which, however, does
not correspond to achiral configuations. We could not find an analytic
expression for this line which emerges from $\f \approx 0.46 \p$, $h/\r=0$ and
ends in $ \f \approx 0.88 \p$, $h/\r=0$.  The three node lines of the chirality
index meet in a single point which corresponds to a regular tetrahedron. For
the fact that a vanishing value of the chirality index does not necessarily
correspond to an achiral configuration,  see in Ref.~\cite{HKL}.

\section{Results}
\label{III}

For our investigations, we selected spiral molecule configurations on four cuts
in the parameter plane displayed in Fig.~\ref{fig2}(a).  Three of these cuts
(A, B, C) are taken at constant relative rise $h/\r=0.2073, 0.584$, and
$2.5162$, respectively, each one for $\f \in (0,\p)$.  For the two smaller
values of $h/\r$ we chose a relatively large radius $\r =7$ in order to avoid
any collision of the atoms within a molecule, and the smaller radius $\r =2$
for the large relative rise. The vertical cut D is determined by the screw angle $\f=2\p/3$, radius $\r=3$ and relative rise $h/\r \in (0,5)$.

\begin{figure}
\begin{center}
\includegraphics[scale=1.2]{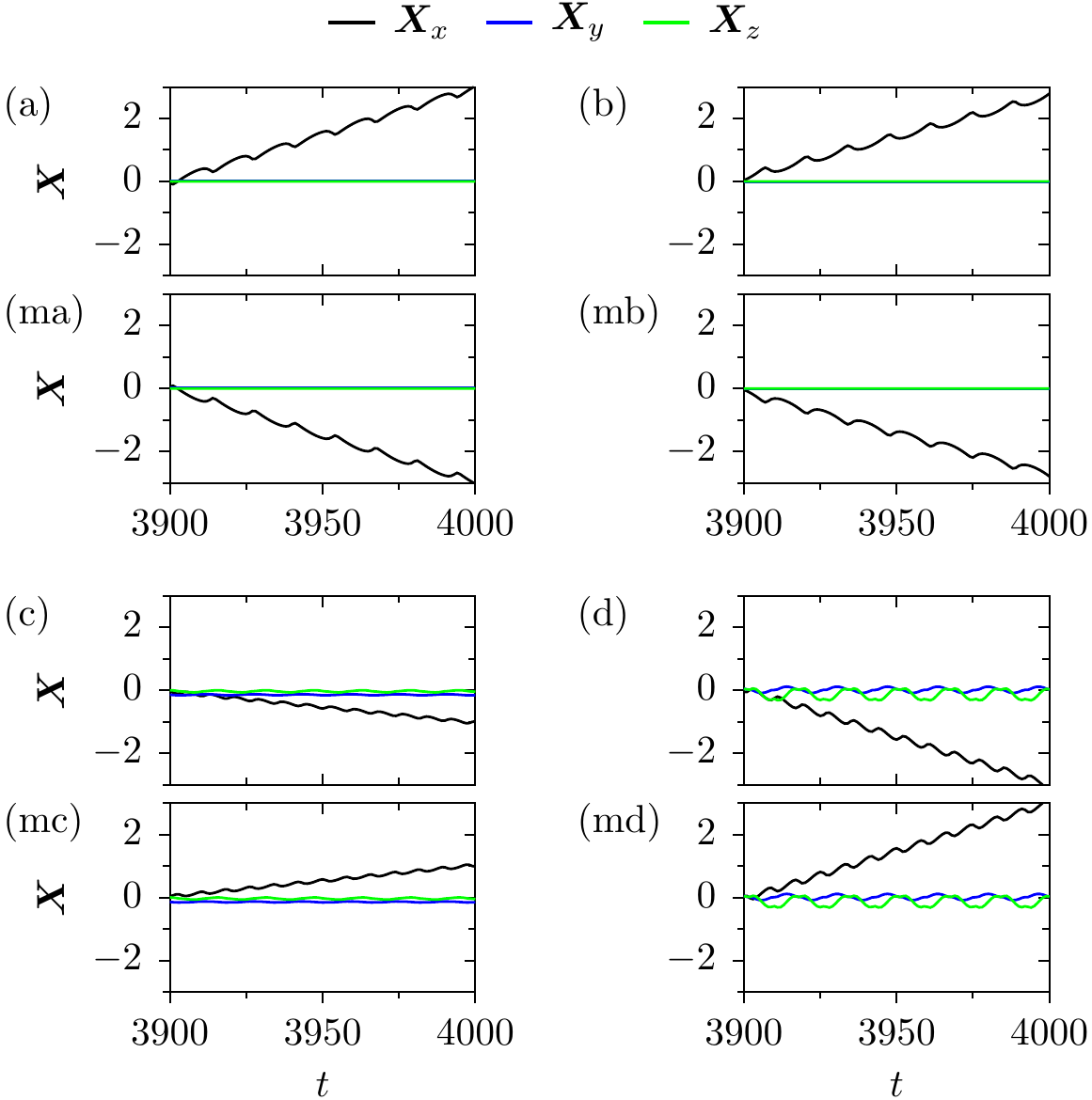}
\end{center}
\caption{The center of mass motion of a spiral molecule with  an equilibrium
configuration on cut A specified by $h/\r = 0.2073$, $\r =7$ and  $\f=0.87 \p$
asymptotically reaches one of the four different states depicted in the panels
(a), (b), (c), (d) depending on the initial orientation of the molecule. If
this orientation is randomly drawn from a uniform distribution, then one finds
the cases (a) with a probability of $0.19$, (b) with $0.22$, (c) with $0.50$,
and (d) with $0.09$.  The $x$-component of the center of mass is displayed in
black. It is shifted by its actual value taken at the time t=3900; the
$y$-component, which is either constant, or periodically oscillating is
depicted by the blue line; from the $z$-component (green) an average uniform
motion is subtracted.  The configurations mirrored at the $y$-$z$-plane
starting from the  corresponding mirror images of the initial orientations of
(a)-(d) result in the motion patterns displayed in the panels (ma)-(md), which
obviously coincide with the mirror images of panels (a)-(d).  The average
velocities $\langle v_x \rangle$ in $x$-direction, see eq.~(\ref{avvx}) below,
in the cases (a) and (ma) are  $\pm 0.02974$;  correspondingly one finds
$\langle v_x \rangle = \pm 0.02594$ in (b) and (mb),   $\langle v_x \rangle =
\mp 0.00994$ in (c) and (mc), and  $\langle v_x \rangle = \mp 0.03044 $ in (d)
and (md).  In all cases the mirrored patterns occur with the probabilities of
the corresponding original ones as specified above.}
\label{fig4}
\end{figure}

As initial configurations we took the equilibrium configurations according to
eq.~(\ref{xn}), rigidly translated such that the center of mass falls into the
origin. Finally,  rigid rotations of the molecule about its center of mass were
performed. Between 500 and 5000 rotations were sampled from the uniform
distribution on the three-dimensional rotation group \cite{rotation}, i.e. from
the Haar measure of SO(3). We typically let run each trajectory for 4000 time
units [$\kappa^{-1}$]. In the majority of the cases the trajectories  have
relaxed toward a stationary regime long before the time span of $4000$.  In
this state the center of mass acquires a finite velocity in $z$-direction,
depending on the precise initial position of the molecule, and the
$y$-component of the velocity may perform small oscillations but vanishes on
average.  The lateral motion in the direction of the vorticity coinciding with
the  $x$-direction in the present case is characterized by an average velocity
superimposed by bounded modulations which were always periodic.
Fig.~\ref{fig4}  displays the components of the center of mass for the same
molecule and its mirror image in four different asymptotic motional states.

\subsection{Center of mass motion}
\label{IIIA}
The fact that the $x$-component of the center of mass moves at a finite average
velocity is apparently related to the chirality of the considered molecule: For
any achiral molecule in the considered shear flow the opposite $x$-directions
are equivalent. This symmetry is broken by the presence of a  chiral molecule
and, according to the Curie principle \cite{C}, the molecule will perform a
directed motion in either of the two directions. This argument is further
corroborated if one launches the chiral partner of a considered molecule into
the shear flow with initial atomic positions obtained from the original ones by
mirroring on the $y$-$z$ plane. Then one indeed finds the molecule moving in
the opposite $x$-direction as illustrated by the panel pairs ((a),(ma)),
((b),(mb)), etc. in Fig.~\ref{fig4} displaying the corresponding pairs of mirror images.

\begin{figure}
\begin{center}
\includegraphics[scale=1.2]{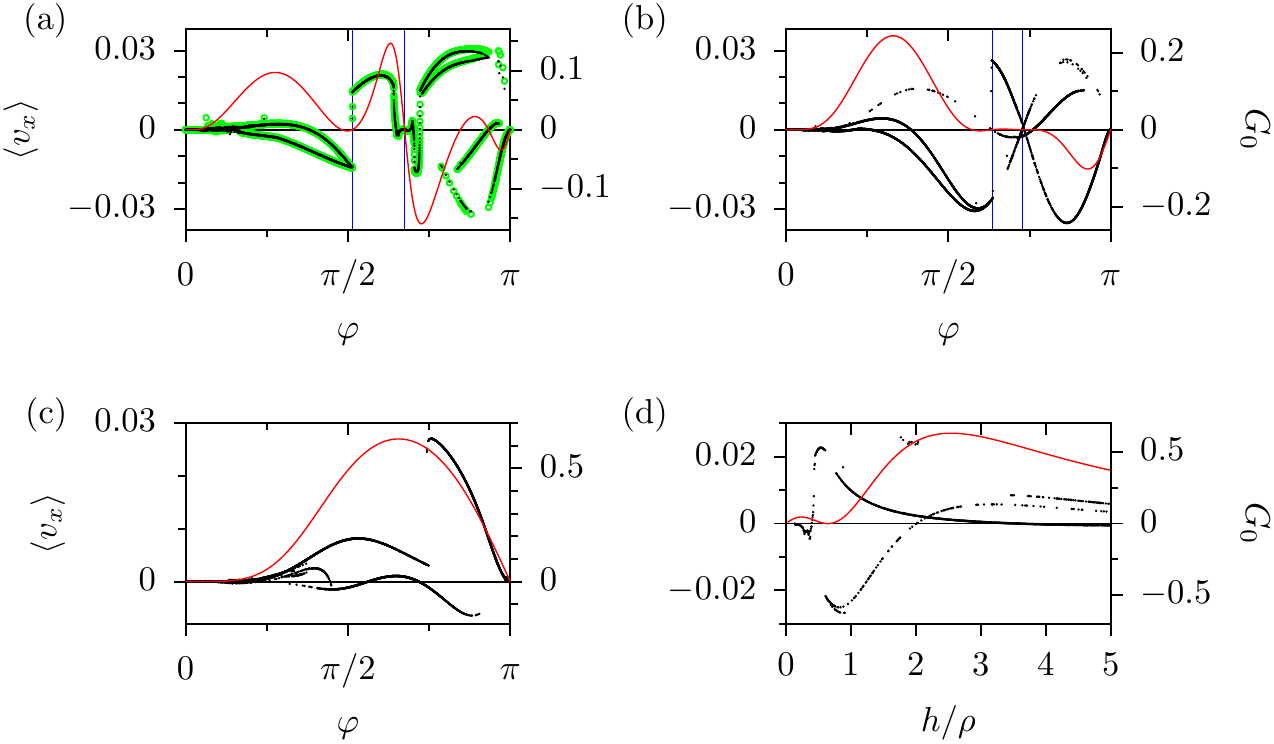}
\end{center}
\caption{Asymptotic average velocities $\langle v_x \rangle$ were estimated for
molecules with equilibrium configurations along the four cuts A -- D in the
$\f$-$h/\r$ parameter plane displayed in Fig.~\ref{fig2}(a). The initial
conditions were randomly oriented according to the uniform measure on the
three-dimensional rotation group. Panels (a), (b) and (c) correspond to the
horizontal cuts A, B, C, along $h/\r=0.2073$ with $\r=7$,  $h/\r= 0.584$ with
$\r =7$, as well as $h/\r= 2.5162$ with $\r=3$, respectively. Panel (d)
represents the vertical cut  D for $\f =2\p/3$, $\r =5$, and $h/\r \in (0,5)$.
For comparison, the chirality index $G_0$ defined in eq.~(\ref{G0}) is
indicated as a red line with the corresponding scale at the right-hand side
ordinate. The vertical blue lines in panels (a) and (b) mark the respective
values of the screw angles at which the equilibrium configurations are achiral.
In all panels the black dots represent the result of the solution of the
equations of motion~(\ref{del}) with the full Rotne-Prager tensor (\ref{RP}).
The average velocities group into up to four well defined branches at most
parameter values. In panel (a), additionally the numerical results obtained
from eq.~(\ref{del}) with the Oseen tensor (\ref{HOseen}) are displayed as
small green circles. The overall agreement of the two approaches is very good,
see also Section~\ref{IV}.  }
\label{fig5}
\end{figure}

Inserting a molecule into the shear flow with randomly chosen orientations  as
described above, one typically finds between one and four different motional
states of the center of mass, which are asymptotically reached at large times.
We determined the corresponding average velocities $\langle v_x \rangle$ by a
least-square fit of the $x$-components of the center of mass to a uniform
motion, yielding
\be
\langle v_x \rangle =  \frac{12}{N(N+1)(N+2) \t} \sum_{n=0}^N (n-\frac{N}{2})X_x(t_0+n \t)\:.
\ee{avvx}
In most cases we used $t_0=3000$, $N=1000$ and $\t =1$.  Since the different
motional states typically are distinguished by their average velocities, these
velocities may be used to roughly classify the asymptotic states as exemplified
in Fig.~\ref{fig5}. A precise classification would require the additional
knowledge of the molecule's orientation and six bond lengths and hence would be
difficult to visualize.

In addition to the average center of mass velocity  the behavior of the
chirality index $G_0$ of the initial configuration is depicted as red line in
Fig.~\ref{fig5} along the respective cuts. This index correlates with the
average velocities at best in a qualitative way. Of particular interest here is
the behavior of the average velocity in the vicinities of the zeroes of the
chirality index as a function of the screw angle $\f$ indicated by the blue
lines.

As mentioned at the end of Section~\ref{II}, one has to distinguish between
chiral and achiral zeroes of the chirality index $G_0$. Indeed, close to chiral
zeroes, none of the there existing velocity branches does cross the zero line.
But also near the achiral zeroes, branches with finite velocities may exist.
They may coexist with  a branch with vanishing average velocity, which,
however, may also be missing.  The $x\leftrightarrow -x$-inversion symmetry
though is restored on average since there are always pairs of branches with
opposite velocities which are populated with equal probability. As we shall
discuss in more detail below, the branches with finite average velocities at
achiral zeroes of the chirality index represent states of dynamically broken
symmetry.

\subsection{Rotational motion}
\label{IIIB}
\begin{figure}
\begin{center}
\includegraphics[scale=1.2]{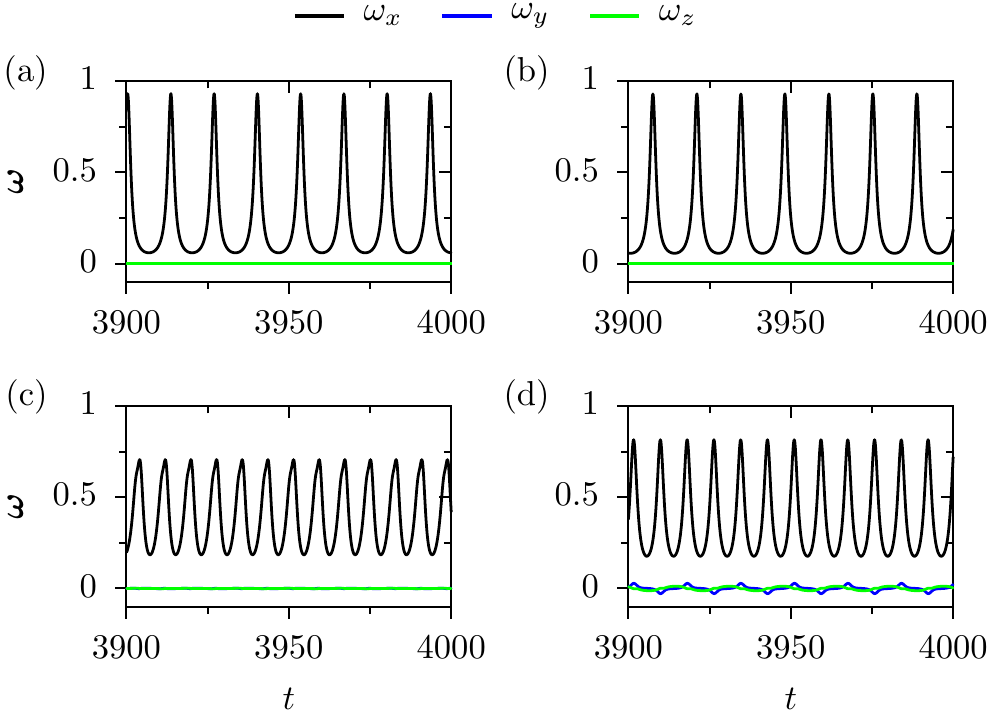}
\end{center}
\caption{The components of the instantaneous axis of rotation, $\bfo$, for the
configuration with $\f=0.87 \p$, $h/\r = 0.2073$ and $\r =7$ corresponding to the maximum of $G_0$ on cut A. The component
$\om_x$ pointing in the vorticity direction of the shear flow is marked in
black, the $y$- and $z$-components in blue and green, respectively. Panels are
labeled as in Fig.~\ref{fig4}. In all cases the rotation about the vorticity
direction is dominant. Only in panel (d) a slight tumbling of the axis is
noticeable. }
\label{fig6}
\end{figure}

\begin{figure}
\begin{center}
\includegraphics[scale=1.2]{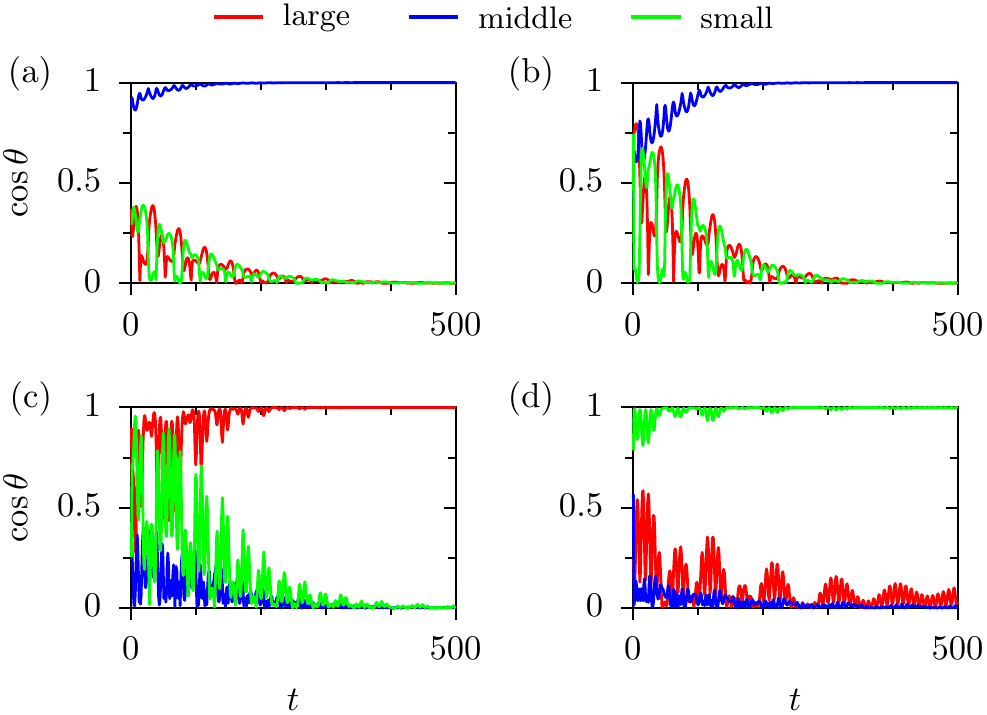}
\end{center}
\caption{The time evolution of the cosines of the angles between the
instantaneous vector of rotation and the eigenvectors of the second moments of
positions, $\cos \theta_i = \bfo \cdot \bu_i/|\bfo|$, for the configuration
with $\f=0.87 \p$, $h/\r = 0.2073$ and $\r =7$ is depicted for different
initial orientations. The asymptotic motion of the center of mass as well as
the motion of the axes of instantaneous rotations are displayed in Figs.
\ref{fig4} and \ref{fig6}, respectively, with the panels labeled consistently.
The cosine of the eigenvector corresponding to the largest eigenvalue is red,
to the middle one blue and to the smallest one green. In panels (a) and (b) the
middle principal axis asymptotically aligns with the rotation axis, while the
initial orientations in panels (c) and (d) lead to alignments with the largest
and the shortest axis, respectively.}
\label{fig7}
\end{figure}

In the asymptotic state, when the center of mass undergoes a periodically
modulated uniform motion, the instantaneous rotation axis predominantly  points
into the $x$-direction, possibly superimposed by small, oscillatory deviations
orthogonal to this direction. Fig.~\ref{fig6} depicts the  components of the
instantaneous rotation axis for different asymptotic motional states of the
same molecule. This orientation of the instantaneous rotation axis appears as a rather natural consequence  of the considered
flow geometry with uniform vorticity pointing into the $x$-direction. A
conspicuous tumbling behavior of the instantaneous rotation axis was found in a
case of dynamically broken symmetry, see Fig~\ref{fig12}a. The rotation speed
undergoes pronounced  oscillations in the majority of cases.  In the asymptotic
states in which the instantaneous rotation axis points in the $x$-direction the
molecule is typically found to be oriented such that one of its principal axes
is aligned with the direction of rotation. The transient behavior and the
approach to the specific asymptotic states of the angles enclosed by the
principal axes and the instantaneous rotation axis is exemplified in
Fig.~\ref{fig7}.

\begin{figure}
\begin{center}
\includegraphics[scale=1.2]{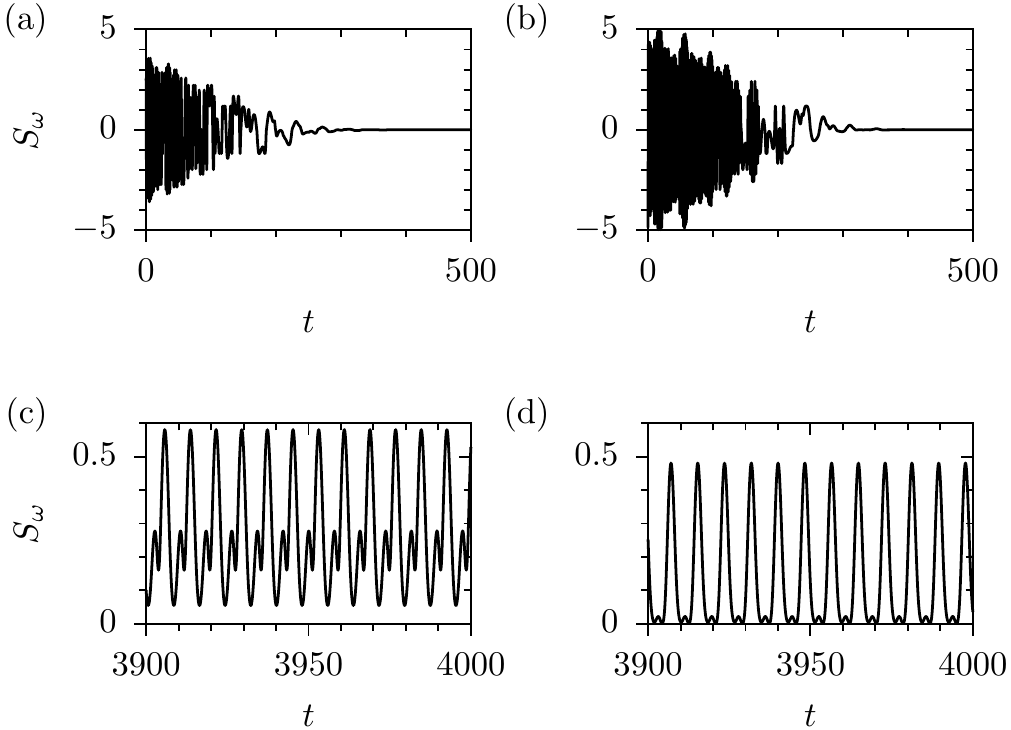}
\end{center}
\caption{For those initial orientations of the configuration with $\f=0.87 \p$
$h/\r = 0.2073$ and $\r =7$ which finally approach two different states both
rotating about the middle axis, see Fig.~\ref{fig7} (a), (b), the third moments
of the positions projected onto the instantaneous axis of rotation, $S_{\bfo}$,
defined in eq.~(\ref{So}),  relax to zero. For other initial orientations
the skewness $S_{\bfo}$ asymptotically undergoes rotations, see panels (c) and
(d).}
\label{fig8}
\end{figure}

Due to the lack of any symmetry plane in a chiral molecule, there are two
possible ways to orient a principal axis along the instantaneous axis of
rotation.  However, we  found different velocity branches corresponding to
opposite orientations only for alignments of the rotation axis with the middle
principal axis, and only if the skewness $S_{\bfo}$ of the molecule along the
axis of rotation vanishes within numerical precision. As a measure for the
skewness of a molecule  relative to the vector of angular velocity $\bfo$ we
take the third root of the third moment of the atomic positions $\bx_n$
relative to the center of mass projected onto the direction of the rotation
axis, $\bfe_{\bfo} = \bfo /|\bfo|$, hence, reading
\be
S_{\bfo} = \left (\sum_n \Big ( (\bx_n -\bX)\cdot \bfe_{\bfo}\Big )^3 \right)^{1/3} \:.
\ee{So}
The maximum number of branches we detected was four. For $\h/\r = 0.2073$, $\r
=7$ and screw angles approximately ranging between $\f=0.8 \p$ and $\f= 0.9 \p$
one observes single alignments to the small and the large axes and both
orientations relative to the middle axis.  Fig.~\ref{fig8} represents the time
evolution of the skewness for a molecule with initial orientations belonging
to different domains of attraction.

\subsection{Internal motion and dynamical symmetry breaking}
\label{IIIC}
\begin{figure}
\begin{center}
\includegraphics[scale=1.2]{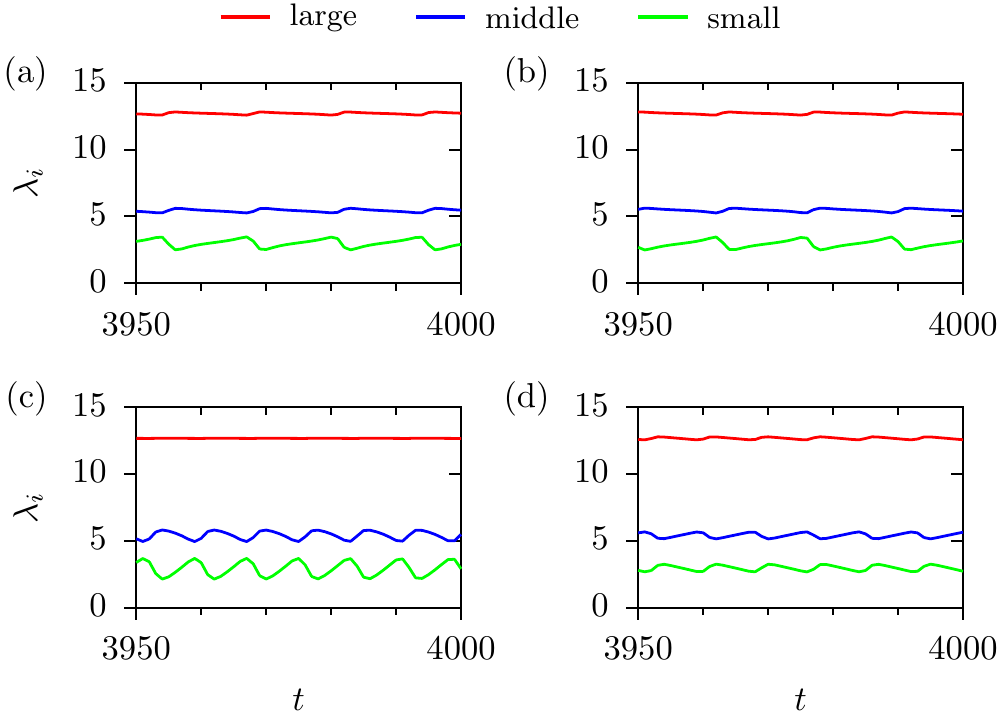}
\end{center}
\caption{The magnitude of the principal axes of the second position moments for
the configuration with $\f=0.87 \p$, $h/\r = 0.2073$ and $\r =7$. In the four
different asymptotic states, which are labeled in the same way as in the
previous Figures, the magnitudes of the principal axes undergo periodic
modifications. In the two cases with positive asymptotic velocities, (a) and
(b), these modifications are almost identical. Noticeable differences occur
between the states characterized by rotations about the large and the short
principal axes, displayed in panels (c) and (d), respectively.}
\label{fig9}
\end{figure}

\begin{figure}
\begin{center}
\includegraphics[scale=1.2]{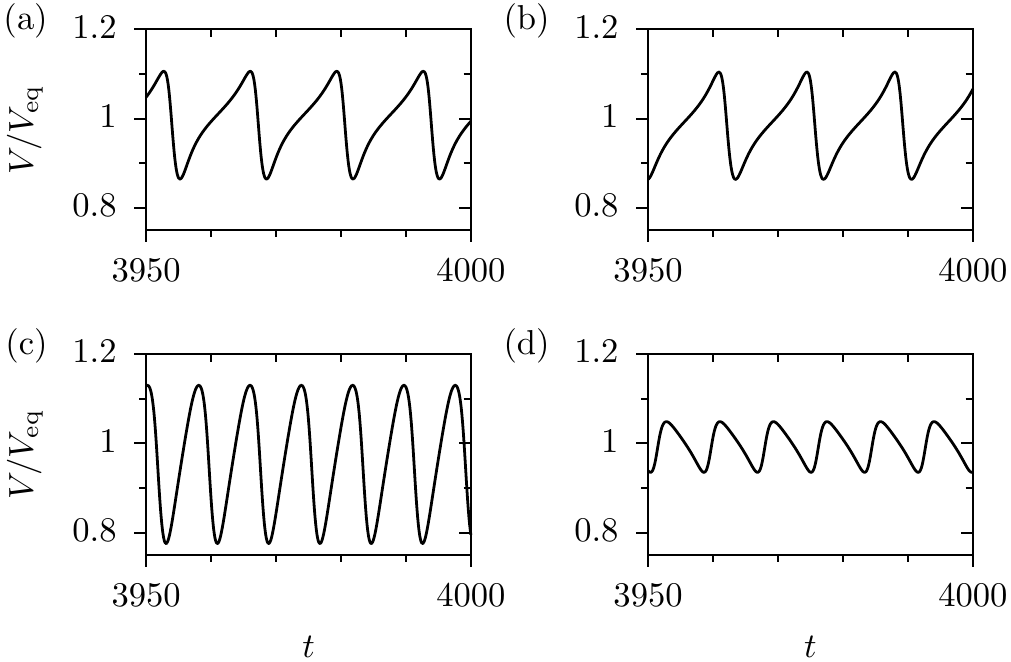}
\end{center}
\caption{The ratio of the actual to the equilibrium volume for different
initial orientations of the configuration with $\f=0.87 \p$, $h/\r = 0.2073$
and $\r =7$ in the same order as in Figs.~\ref{fig6} -- \ref{fig9}. For the two asymptotic states (a) and (b) moving in the positive
$x$-direction the volume variations are essentially identical but different
from both states (c) and (d).}
\label{fig10}
\end{figure}

Also in the reference frame that moves with the center of mass and rotates with
the instantaneous angular velocity, the molecules generally  are not at rest
but undergo internal motions characterized by changing bond lengths and bond
angles. These distortions  then lead to modulations of the magnitudes of the
principal axes as exemplified in Fig.~\ref{fig9}. The resulting periodic
relative volume changes of the molecule are presented in Fig.~\ref{fig10}. 

\subsubsection{Dynamical symmetry breaking}
\label{IIIC1}
In molecules with achiral equilibrium configurations, a deformation caused by
shear forces in general will break its mirror symmetry  and hence render the
molecules  chiral. This mechanism is at the heart of the effect of {\it
dynamical symmetry breaking} that can be observed at  achiral zeroes of the
chirality index $G_0$.

\begin{figure}
\begin{center}
\includegraphics[width=0.8\columnwidth]{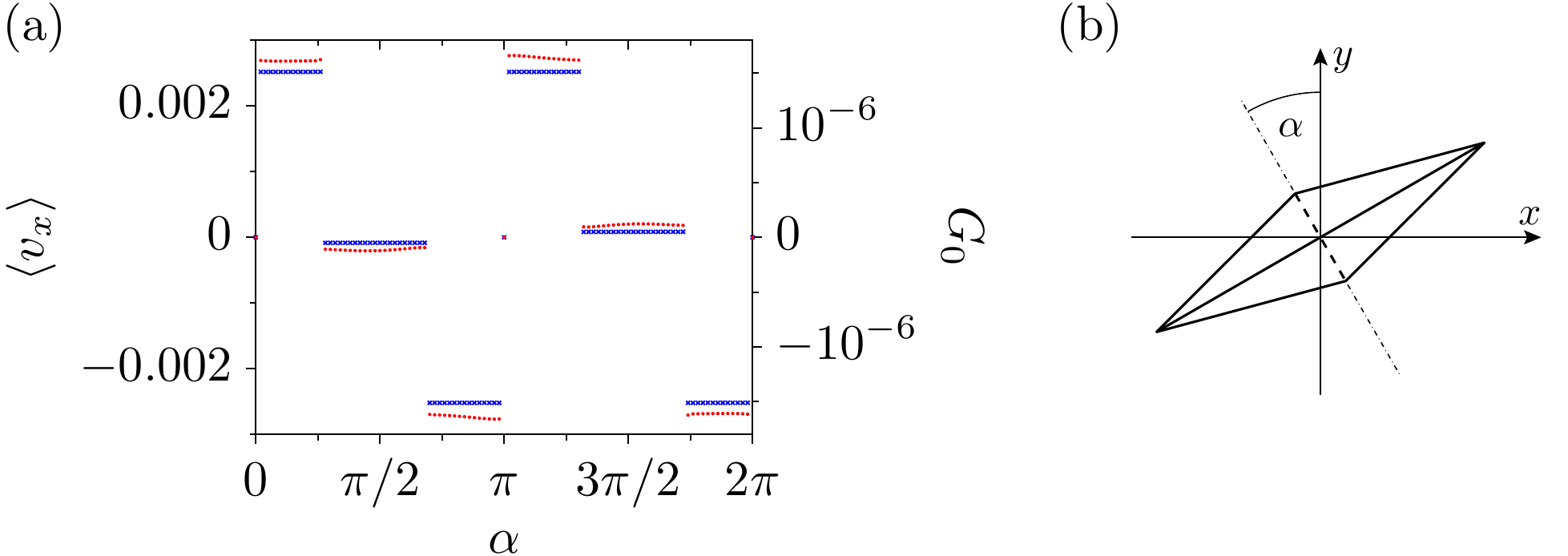}
\end{center}
\caption{On the panel (a), different chiral-symmetry breaking attractors for
the achiral equilibrium configuration at $\f = 0.7271607 \p$, $h/\r = 0.584$ and
$\r=7$ are characterized by their average velocities $\langle v_x \rangle$
(blue) and average chiral indices  $\langle G_0 \rangle $ (red). The attractors
are approached from different orientations of the molecule. The angle $\a$
specifying the initial orientation is enclosed by the $y$ axis and the
molecule's symmetry plane containing the short molecular edge as exemplified in
the panel (b) which presents a view with the $z$-axis perpendicular to the
plane of projection. For $\a= 0$ the $y$-$z$-plane coincides with this symmetry
plane of the molecule.}
\label{fig11}
\end{figure}

\begin{figure}
\begin{center}
\includegraphics[scale=1.2]{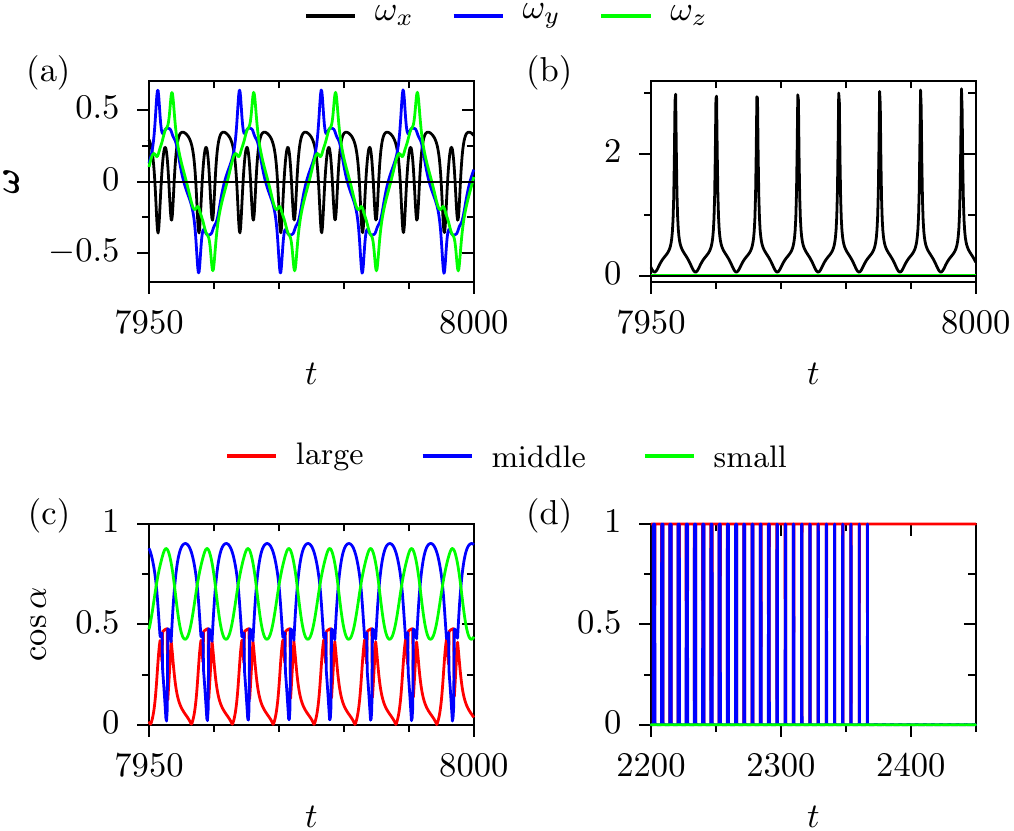}
\end{center}
\caption{Different modes of the rotational motion of a molecule with $\f
=0.7271607 \p $, $h/\r= 0.584$ and $\r=7$ corresponding to an achiral
equilibrium configuration on the $(h/\r)_1$-curve. In the upper row the components $\omega_x$ (black),
$\omega_y$ (blue), and  $\omega_z$ (green)  of the instantaneous rotation axis
are displayed; the lower row presents the cosines of angles between the large
(red), middle (blue) and small (green) principal axes and the $x$-axis. The
left column corresponds to the asymptotic state with larger positive velocity
in $x$-direction $\langle v_x \rangle = 0.0025$ and the right column to the asymptotic state with smaller
positive velocity in $x$-direction, $\langle v_x \rangle =0.000083$. In case of the large positive velocity the
molecule performs a tumbling but yet periodic motion as can be seen both from
the oscillatory behavior of the instantaneous rotation axis (a) and the
alignment of the instantaneous axis with the $x$-axis (c). In the
asymptotic state of slow motion the instantaneous rotation axis exactly points
into the $x$-direction (b). Asymptotically (for times $t> 2360$) the large
principal axis strictly lies in the $x$-direction (d). In a transient period at
earlier times the molecular orientation switches between perfect $x$-alignments
with the large and the middle principal axes (d).}
\label{fig12}
\end{figure}

For spiral molecules the three-dimensional achiral configurations are found on
two curves in the $\f$-$h/\r$-parameter plane, reading
\ba
\label{acp1}
(h/\r)_1& =\left [\frac{2}{3} (\cos (2 \f) - \cos (\f) ) \right]^{1/2}, \quad 2 \p/3 < \f < 4 \p /3\:,\\
(h/\r)_2&= \frac{1}{2} \left [(\cos (3 \f) - \cos (\f) ) \right]^{1/2}, \quad \p/2 < \f < 3\p/2 \: .
\ea{acp2}
For a derivation see the Appendix. The branch $(h/\r)_1$ corresponds to the
solid line and $(h/\r)_2$ to the dashed line in the Fig.~\ref{fig2}(b).
Fig.~\ref{fig11}(a) displays the asymptotic average velocities in $x$-direction
and the corresponding average values of the chirality index for the achiral
configuration at $\f=0.7271607 \p$ and $\rho=7$ on the curve $(h/\r)_1$. As for
all achiral configurations on this curve, it possess five equally long edges.
In the present case, the sixth edge is shorter than the others. This molecule
is initially  oriented in such a way that the symmetry plane  containing both
the short edge and the $z$-axis encloses an angle $\a$ with the $y$-axis, see
the Fig~\ref{fig11}(b).  Depending on this angle $\a$, two different  pairs of
asymptotic  states  with non-vanishing velocities and chiral indices of
opposite signs are approached with time. The average chirality index was
obtained as an algebraic mean over a time window of length 1000 in the
asymptotic regime.  Because a rotation about the $z$-axis by $\a = \p$ leaves
any of these initial configurations invariant the resulting pattern of average
velocities and chiral indices is periodic with period $\p$.  The rotational
motion and the orientation of the principal axes are presented in
Fig.~\ref{fig12} for two of the chiral attractors of an achiral molecule.

\begin{figure}
\begin{center}
\includegraphics[scale=1.2]{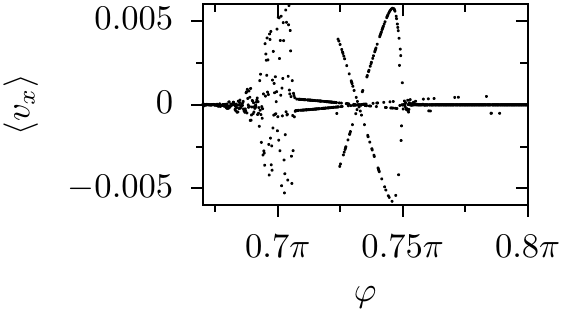}
\end{center}
\caption{Average velocities in the $x$-direction for 2000 randomly rotated
initial conditions taken along the achiral curve $(h/\r)_1$ given  by eq.~(\ref{acp1}) are
displayed for $\r=5$ and $\f \in [0.67 \pi, 0.8 \p]$ in steps of $\Delta \f =
1.65 \cdot 10^{-4}\:\pi$. The equations of motion~(\ref{del}) were numerically
solved with these initial conditions up to the time $t=8000$ and the average
$\langle v_x\rangle$ was estimated by means of eq.~(\ref{avvx}) for $t_0=7000$,
$\t =1$ and $N=1000$. Apart from the region $\f \in (0.68 \p, 0.7066 \pi)$
where the average velocities seemingly are random, they form up to four
branches. Coming in pairs of opposite average velocities they  maintain the
achiral symmetry at each point along the achiral curve. On the part of the
achiral curve with $\f \in [0.8 \p, \p]$, which is not shown, the lateral
average velocity forms a single branch with $\langle v_x \rangle =0$ within
numerical precision.}
\label{fig13}
\end{figure}

\begin{figure}
\begin{center}
\includegraphics[scale=1.2]{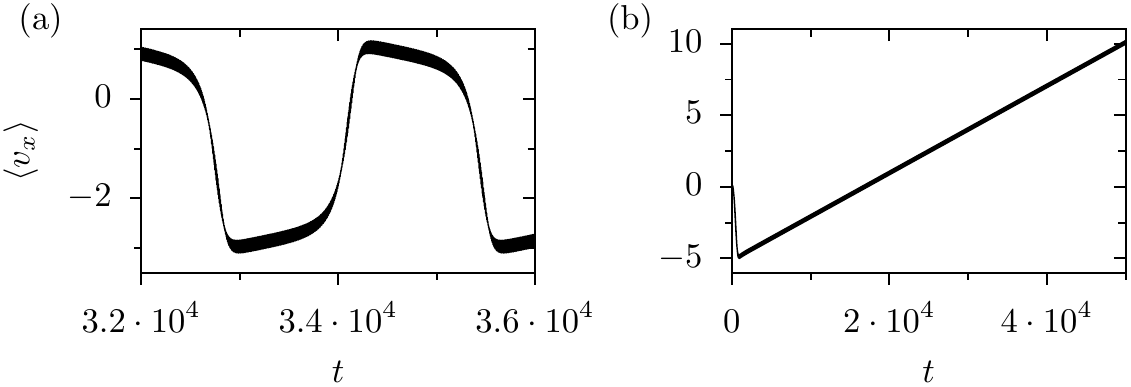}
\end{center}
\caption{The center of mass motion on the achiral curve, eq. (\ref{acp1}) (a)
for $\f=0.7 \p$, and (b) $\f = 0.71092 \p$, both for $\r=5$. The long period
$T\approx 2730$ in (a) leads to an under-sampling of the average $\langle v_x
\rangle$ displayed in Fig.~\ref{fig13} and as a consequence to the
cloud of random points in the screw angle region between $0.66 \p$ and
$0.7066 \p$. In contrast, the case in panel (b) approaches a uniform
motion in the positive $x$ direction after a short transient. }
\label{fig14}
\end{figure}

The average asymptotic velocities on a part of the achiral curve $(h/\r)_1$
given by eq.~(\ref{acp1}) %are displayed in Fig.~\ref{fig13}  
were determined for $\r=5$.
Similarly as on the other cuts, equilibrium configurations were determined from
eq.~(\ref{xn}) at 2001 equally spaced angular values for $\f \in [0.67 \p,\p]$;
subsequently their centers of mass were translated to the origin and randomly
rotated about the center of mass. With these initial conditions the  equations
of motion were integrated up to $t= 8000$. The average velocity $\langle v_x
\rangle$ was determined according to eq.~(\ref{avvx}) for $t_0=7000$, $N=1000$
and $\t=1$.  Fig.~\ref{fig13} presents the resulting averaged velocities in
$x$-direction. The emerging pattern is essentially symmetric about $\langle v_x
\rangle =0$ and consists  of different branches and a region with a cloud of
apparently random points. The deviations from symmetry are insignificant within
statistical uncertainty.  While, on the branches, the center of mass approaches
an average motion with constant velocity $\langle v_x \rangle$, superimposed by
short oscillations, in the cloud region it asymptotically oscillates with a
period which is larger than the width of the averaging window, see in
Fig.~\ref{fig14}. Hence, the resulting averages over a window that is shorter than a period are randomly distributed
within an upper and a lower bound. The true averages  extended over a full
period are zero within  numerical precision.

As for the case of chiral equilibrium configurations, also for achiral
equilibrium configurations the number of branches in the present investigation
varies from one up to four.  In the limiting cases of $\f = 2/3 \p$ and $\f=\p$
the configurations on the achiral curve are planar and only a single branch
with $\langle v_x \rangle = 0$ exists. Additional branches appear for
intermediate values of $\f$ always in pairs with positive and negative
velocities and opposite chiral indices, while the zero-velocity and
zero-chirality-index branch may become unstable. Branches with opposite
velocities and chiral indices always occur with equal weight such that the
mirror symmetry of the initial achiral configuration is restored.  For $\f =
\arccos (-2/3) \approx 0.732 \p$, corresponding to a regular tetrahedron, four
velocity branches cross each other at $\langle v_x \rangle =0$, see
Fig.~\ref{fig13}.

The present phenomenon of dynamical symmetry breaking is different from
the one that was reported in Ref.~\cite{WL} where molecules with initially
achiral geometric configurations, but with different bond strengths, were
considered. Hence those  molecules were chiral from the outset  with respect to
their mechanical properties. We emphasize here that the  effect of dynamical
symmetry breaking occurs for achiral molecules with respect to both geometry
and bond strengths.

\section{Role of hydrodynamic interactions}
\label{IV}
\begin{figure}
\begin{center}
\includegraphics[scale=1.2]{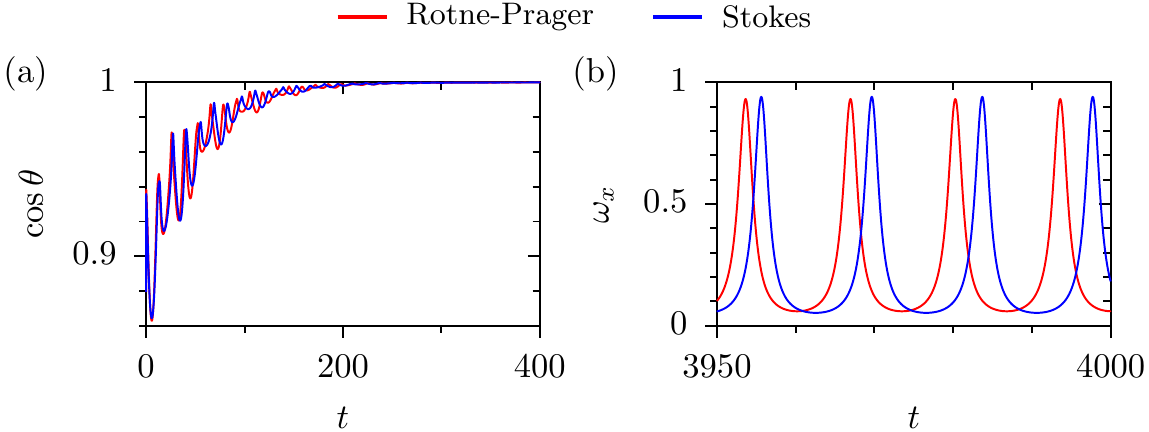}
\end{center}
\caption{A comparison of the rotational motion of a molecule under the sole
action of the Stokes forces (blue) and the full Rotne-Prager tensor (red) is
presented for a spiral molecule with $\f= 0.87$, $h/\r= 0.2073$ and $\r=7$ with
an initial orientation that leads to the uppermost velocity branch (see
Fig.~\ref{fig5}(a)) within the Rotne-Prager treatment. The panel (a) displays
the approach of the angle between the instantaneous rotation axis and the
middle principal axis, showing good agreement of the Stokes and Rotne-Prager
treatment. (b) The oscillations of the $x$-component of the instantaneous rotation
axis follow the same pattern with the same amplitude and slightly smaller
period under the influence of the Rotne-Prager mobility as for the Stokes
forces.}
\label{fig15}
\end{figure}

In order to clarify the role of hydrodynamic interactions for the transport of
chiral objects in a shear flow, we compared the motion of a spiral molecule
with screw angle $\f =\p/2$, relative rise $h/\r =0.584$ and radius $\r=7$ in
presence of three different vector fields describing the action of the fluid on
the molecule at different levels of accuracy. In the crudest approximation the
forces contribute to the velocity in terms of the Stokes law for which the
mobility tensor simplifies to
\be
\bH^{\mathrm{Stokes}}_{i,j} = \frac{1}{6 \p \h a} \mathbbm{1} \d_{i,j} \;.
\ee{HStokes}
Leading corrections in the ratios of the atomic radius to intra-molecular
distances are contained within the Oseen approximation, yielding
\be
\bH^{\mathrm{Oseen}}_{i,j} = \bH^{\mathrm{Stokes}}_{i,j} +\frac{1-\d_{i,j}}{8 \p \h \br_{i,j}} \left [ \mathbbm{1} +\frac{\br_{i,j} \br_{i,j}}{r^2_{i,j}} \right ] \;.
\ee{HOseen}
Finally we compared the Stokes and Oseen results  with those following from the
full Rotne-Prager tensor (\ref{RP}). Since we found only minor differences
between the outcomes of the Oseen and Rotne-Prager treatment, see Fig.~\ref{fig5}(a),
we only present results of the comparison between Stokes and Rotne-Prager. It is interesting to note that the
rotational motion and the internal motion of the molecule qualitatively are
quite similar for the Stokes and the Rotne-Prager treatment. 
Fig.~\ref{fig15} exemplifies the qualitative agreement of the two approximations by comparing the alignment of (a) the middle principal axis with the
instantaneous rotation axis and (b) the $x$-component of the instantaneous
rotation axis.

\begin{figure}
\begin{center}
\includegraphics[scale=1.2]{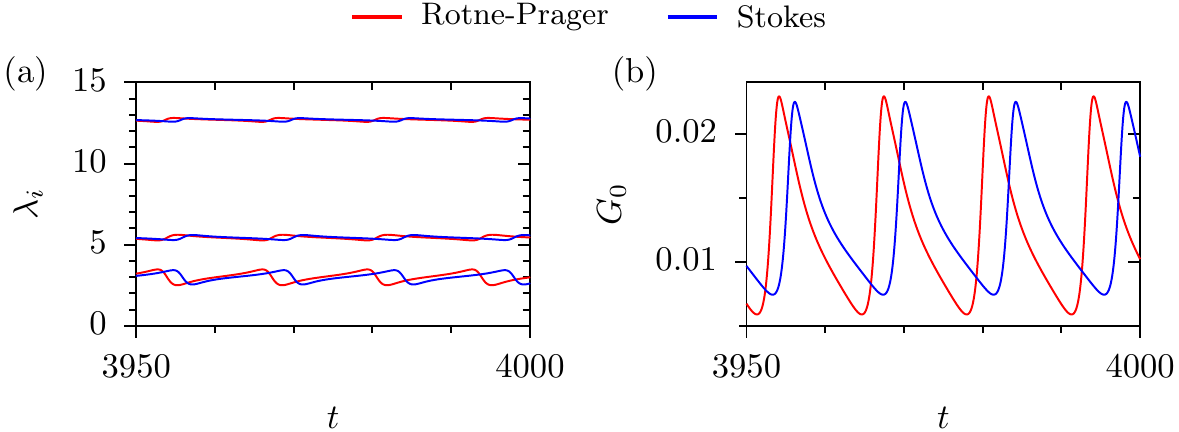}
\end{center}
\caption{For the parameter values detailed in Fig.~\ref{fig15}, the
hydrodynamic interactions contained in the full Rotne-Prager mobility tensor
(red) have little influence on the magnitudes of the principal axes displayed
in panel (a) compared to the sole action of Stokes forces (blue). Also the
chirality index presented in (b) differs in a small shift of the period, which
is slightly shorter, and in the amplitude which is larger in the case of the
Rotne-Prager treatment.}
\label{fig16}
\end{figure}

Also the deformational motion of the molecule is quite similar, see
Fig.~\ref{fig16} comparing the magnitudes of the principal axes and the
relative volume for the two approximations. Hence the rotational and internal
motion of a molecule is mainly determined by the action of the Stokes forces.

In the present case of uniform shear flow the sole action of the
Stokes forces cannot lead to a translational motion in the vorticity
direction.
This rigorously follows from the fact that the internal molecular
forces exerted by the intra-molecular FENE-Fraenkel springs  do not contribute
to the motion of the center of mass for a uniform shear flow within the Stokes
approximation, $\sum_{i,j} \bH^{\mathrm{Stokes}}_{i,j} \bF_j =0$, and hence the
center of mass motion is determined by
\be
\dot{\bX} = \sum_n \bv(\bx_n)\:.
\ee{cmS}
Therefore, in a uniform shear flow, as given by eq.~(\ref{shv}), the center of
mass cannot move perpendicularly to the flow direction within the Stokes
approximation and the lateral drift consequently vanishes.  
Only the small, non-diagonal terms of the mobility tensor implied by both
the Oseen and Rotne-Prager treatment lead to a coupling between the
translational and the rotational motion and, in this way facilitate a center of
mass motion perpendicular to the flow direction.
For more complicated flow fields with non-uniform vorticity such a
coupling becomes possible and may lead to chiral separation even within the
Stokes approximation \cite{KSTH}.

\section{Conclusions}
\label{V}
We presented a case study of the dynamics of deformable chiral objects
consisting of four spheres in a uniform shear flow. These objects, which we
have called molecules, consist of four spherical atoms. Internal forces due to
relative position changes of the atoms were modeled by FENE-Fraenkel forces
such that the bond lengths may only vary within a prescribed range. In the
equilibrium configuration the internal forces balance each other. Forces
exerted by the fluid on the molecule drive the molecule out of this equilibrium
state.  Due to the small size and mass of the atoms, inertia terms in the
equations of motion could be neglected and a purely overdamped motion was studied.
The influence of the flowing fluid on the molecules was described by means of
the Rotne-Prager mobility tensor; this tensor includes the Stokes force and
hydrodynamic interactions (up to second order in the atom radius to
intra-molecular distance ratio) between the atoms of a molecule.  If only the
Stokes forces are considered and hydrodynamic interactions are neglected then
the molecule moves in a uniform shear flow with its center of mass strictly in the
direction of the flow field. Moreover it also performs a rotational motion
about the center of mass with an instantaneous rotation axis pointing into the
direction of vorticity.  However, the hydrodynamic interactions cause a
coupling between the translational and the rotational motion of the molecule.
These interactions let the rotating molecule act as a propeller which drives
the molecule  transversely  to the flow field in or against the direction of
its vorticity.  The direction of this component of motion is opposite for a
molecule and its mirror images,  thus leading to different transport properties
of enantiomers.

In general, one finds different asymptotic forms of motion for  molecules with
the same equilibrium configuration.  These motional states correspond to
different attractors of the equations of motion. They can be distinguished by
the magnitude of their average velocities in vorticity direction and the
orientation of the molecule relative to the instantaneous rotation axis which
typically points into the vorticity direction. The maximum number of attractors
that we identified in the present study was four; this corresponds to  (i) two
alignments of the instantaneous axis of rotation with the large and the small
principal axes and (ii) two alignments with the middle axis.
We documented the typical translational, rotational and deformational motion
in detail 
with the example of a molecule with a particular equilibrium configuration. 
The same qualitative motion patterns were observed for numerous other
equilibrium configurations  but not presented here.

The reason why we studied deformable molecules rather than rigid objects was
not only to obtain a more realistic modeling but also because the numerical
implementation of the constraints defining rigid objects in three dimensions is
more difficult than allowing for deformations.  Moreover we found that the
flexibility of the molecules may lead to dynamical symmetry breaking for
molecules with achiral equilibrium configurations. We demonstrated that the
deformations of achiral  molecules may lead to chiral configurations with
finite chirality index $G_0$ accompanied by a finite average lateral velocity
in or opposite to the vorticity direction. If the initial conditions are
sampled from a symmetric distribution, containing mirror-symmetry related
orientations with equal weight, then motional states with opposite chirality
emerge with equal probability. In this way the mirror symmetry of the total
state space is recovered.

The effects of chirality on the transport properties are relatively weak in the
present case of a uniform shear flow. This finding is also in agreement with
the results of Ref.~\cite{WL} for bond-strength induced chirality. The
situation though is expected to drastically change for non-uniform shear flows.
It is plausible to assume that there, locally, still the same principal
mechanisms are at work: The local vorticity leads to a rotation of the molecule
which then  acts via hydrodynamic interactions as a propeller pushing itself in  or
opposite to the direction of the instantaneous rotation axis depending on its
chirality. In non-uniform shear flows molecules of different chirality,
however, will be moved to different regions of the flow-field where they then
are  advected with different velocities. This effect can significantly enhance
the influence of the chirality on transport properties and hence, may form the
basis of an effective, purely physical enantiomer-separation mechanisms.

Before such practical applications can be envisaged, however, the so far
neglected influences of thermal noise, hydrodynamic interactions between
different molecules of same and different chirality as well as between
molecules and walls confining the fluid should all  be taken into account and
investigated.

\ack
We express our sincere thanks to Marcin Kostur and Juyeon Yi for valuable
discussions. This project has been supported by the Deutsche
Forschungsgemeinschaft (DFG), grant number HA-1517/28-1.

\appendix
\label{A}
\section{Achiral configurations}
By the construction of the spiral molecules, the length $d_{n,m}$ of the edge
connecting the $n$th with the $m$th atom only depends on the absolute value of
the difference between $n$ and $m$, $d_{n,m} = l(|n-m|)$.  In the general case,
a spiral molecule hence has three equal edges of length $l(1)$, two of length
$l(2)$ and one of length $l(3)$ where
\be
l(n) = \left [ 2 \r^2 (1-\cos (n \f)) +n^2 h^2 \right ]^{1/2}\:.
\ee{ln}

One branch of achiral configurations possesses two equilateral  and two
isosceles faces, see Fig~\ref{fig3}(c). They have five equally long edges with
$l(1) = l(2)$. This condition determines the relative rise as a function of the
screw angle to read
\be
(h/\r)_1^2 = \frac{2}{3} (\cos (2 \f) -\cos (\f))\;,
\ee{acmc}
where the screw angle is restricted to $2 \p/3 < \f < 4 \p/3$ in order that the
right hand side stays positive.  On this curve the maximal value of the
relative rise is reached at $\f = \p$ and there takes the value $h/\r = 2/\sqrt{3}$. 

Achiral configurations with four identical  isosceles faces result from the
requirement $l(1)=l(3)$. The relative rise as a function of the screw angle
then becomes
\be
(h/\r)_2^2 = \frac{1}{4} (\cos (3 \f) -\cos (\f))\;, 
\ee{acmb}
with $\p/2 < \f < 3\p/2$. This function vanishes at $\f=\pi$ as well as at the borders 
of the indicated interval, and has a maximum at $\f\approx 0.696 \p$, $h/\r\approx 0.620$.

The two achiral branches cross at the special angle at which $\cos \f = -2/3$
yielding $h/\r =\sqrt{10/27}$, where all three edge lengths  agree with the
each other, resulting in a regular tetrahedron.


\begin{thebibliography}{99}
\bibitem{rotation} In order to draw elements out of the Haar measure of SO(3)
we first generate a random vector with three independent Gaussian components
about which a rotation with an angle $\a$ is performed. To guarantee a uniform
coverage of the rotation group, the rotation angle $\a \in [0,\p]$ must be
drawn from the probability density $(1/\p) \sin^2 (\a/2)$, \cite{Miles}.
\end{thebibliography}
\end{document}